\documentclass{vldb}
\usepackage{graphicx}
\usepackage{balance}  

\usepackage[usenames, dvipsnames]{color}
\usepackage{subcaption}
\usepackage{multirow}
\usepackage{hyperref}
\usepackage{enumitem}
\usepackage{booktabs}

\definecolor{d3purple}{rgb}{0.578, 0.402, 0.742}
\definecolor{d3orange}{rgb}{0.999, 0.496, 0.055}

\newdef{definition}{Definition}

\newcommand{\name}[1]{\textit{ETable}}

\begin{document}


\title{Interactive Browsing and Navigation\\in Relational Databases}



%
%
%
%

\numberofauthors{1} 

\author{
%
%
\alignauthor
Minsuk Kahng, Shamkant B. Navathe, John T. Stasko, and Duen Horng (Polo) Chau\\
       \affaddr{Georgia Institute of Technology}\\
       \affaddr{Atlanta, GA, USA}\\
       \email{kahng@gatech.edu, sham@cc.gatech.edu, stasko@cc.gatech.edu, polo@gatech.edu}
}
\date{20 July 2016}

\maketitle

\begin{abstract}

Although researchers have devoted considerable attention to helping database users formulate queries,
many users still find it challenging to specify queries that involve joining tables.
To help users construct join queries for exploring relational databases,
we propose \name{}, a novel presentation data model that provides users with a presentation-level interactive view.
This view compactly presents one-to-many and many-to-many relationships within a single \textit{enriched table} by allowing a cell to contain a set of \textit{entity references}.
Users can directly interact with this enriched table to incrementally construct complex queries and navigate databases on a conceptual entity-relationship level.
In a user study, participants performed a range of database querying tasks faster with \name{} than with a commercial graphical query builder. 
Subjective feedback about \name{} was also positive. All participants found that \name{} was easier to learn and helpful for exploring databases.

\end{abstract}

\section{Introduction}
\label{sec:intro}

\begin{figure*}
  \begin{center}
  \includegraphics[width=\linewidth,trim={7.6cm 17.8cm 6.8cm 2.4cm},clip]{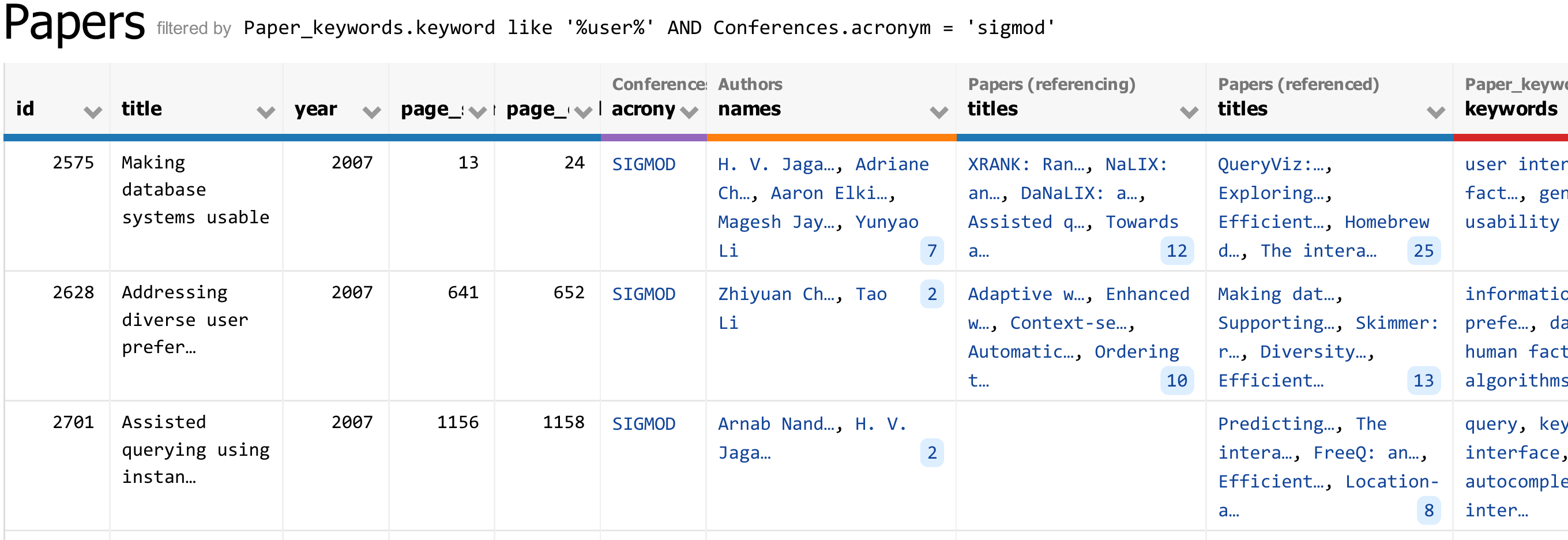}
  \vspace{-18pt}
  \caption{\name{} integrates multiple relations into a single enriched table that helps users browse databases and interactively specify operators for building complex queries. 
  	This example presents a list of SIGMOD papers containing the keyword ``user'' from an academic paper database collected from DBLP and the ACM Digital Library.
  	Each column represents either a base attribute of a paper or a set of relevant entities obtained from other tables (e.g., 
  	\texttt{\textcolor{d3purple}{Conferences}}, \texttt{\textcolor{d3orange}{Authors}}).
  	If a relational database were used to obtain the same information, 9 tables would need to be joined, and the results produced can be hard to interpret because of many duplicated cells.}
  \label{fig:table-screenshot}
  \end{center}
  \vspace{-2pt}
\end{figure*}

A considerable challenge for non-technical users of relational databases is constructing \textit{join} queries~\cite{jagadish2007making}.
The join operation is required for even simple data  lookup queries since relational databases store information in multiple separate normalized tables.
Although database normalization provides many benefits for managing data (e.g., avoiding update anomalies), it significantly decreases the usability of database systems by forcing users to write many join queries to explore databases.

Constructing join queries is difficult for several reasons. 
The main reason is that users find it difficult to determine which relations to join among many relations. Understanding the role of each relation that represents a relationship of interest and finding the right join attributes are not trivial tasks, even when a schema diagram is given.
To tackle this challenge, users often write complex queries by starting with a simpler query and iteratively adding operators~\cite{nandi2011guided}.
Although this iterative strategy is helpful, it is still challenging because the format of join query results is hard to interpret.
For example, consider a query that joins two relations in many-to-many relationships (e.g., \texttt{Papers} and \texttt{Authors} in Figure~\ref{fig:schema}).
A result of this query produces a large number of duplications (e.g., the title of each paper repeated as many times as the number of its authors).
People represent the same information differently when they use a spreadsheet. For instance, they might create a cell containing multiple values separated by commas.
Relational databases cannot represent data in this way because the relational model (as implemented in most relational DBMSs) requires that data be at least in the \textit{first normal form}.

The usability challenge of writing complex queries has been studied by many researchers.
Although \textit{visual query builders} help people formulate SQL queries~\cite{catarci1997visual}, they separate query construction and result presentation parts~\cite{jagadish2007making}, introducing a usability gap between users' actions and their results~\cite{shneiderman1983direct, nandi2011guided}.
To overcome this limitation, researchers argue that database interfaces need to adopt the \textit{direct manipulation} principle~\cite{shneiderman1983direct}, a well-known concept in the \textit{human-computer interaction (HCI)} area~\cite{jagadish2007making, liu2009spreadsheet}.
It enables users to iteratively specify operators by directly interacting with result instances using simple interactions~\cite{liu2009spreadsheet}.
Researchers also argue that join query results should be represented in an easier-to-understand format that improves the interpretation of query results.
Jagadish et al.~\cite{jagadish2011organic} proposed the notion of the \textit{presentation data model}, which they defined as a full-fledged layer above the logical and physical schema.  This presentation layer allows users to better understand the query results without requiring full awareness of the schema.
All this research strongly suggests the need for developing database interfaces that are usable, interactive, and interpretable.

We present \name{}, a novel presentation data model with which users can interactively browse and navigate databases on an entity-relationship level without writing SQL. 
\name{} presents a query result as an enriched table in which each cell can contain a set of \textit{entity references}. 
By deliberately relaxing the \textit{first normal form}, we compactly represent one-to-many and many-to-many relationships within a single table ---
a novel capability that enables users to more easily browse and interpret query results consisting of multiple relations.
Figure~\ref{fig:table-screenshot} illustrates how \name{} effectively presents
a list of SIGMOD papers containing the keyword ``user'' from an academic paper database collected from DBLP and the ACM Digital Library (see Figure~\ref{fig:schema} for schema).
Each row in \name{} shows the base attributes and relevant entities of a paper, such as its authors and cited papers.
If a relational database were used to obtain the same information, 9 tables would need to be joined, and the results produced would be hard to interpret (e.g., many duplicated cells).

To discover which relevant entities should be shown for each row, 
\name{} uses a novel graph-based model called the \textit{typed graph model (TGM)}, which frees users from concerning themselves with the complexity of the logical schema;
users may instead focus on exploring and understanding the data set at the conceptual (or entity-relationship) level.
The \textit{typed graph model} stores relational data as graphs
in which nodes represent entities (e.g., authors, papers) and edges represent relationships (e.g., those that relate authors to papers).
This transformation enables \name{} to retrieve other related entities through simple graph operations.
For example, a given paper's authors, stored as direct neighbors, can be retrieved through a quick neighbor-lookup.

As the construction of complex queries and the exploration of data are inherently iterative processes, 
database exploration tools should provide easy-to-use operations to help users incrementally revise  queries~\cite{cetintemel2013query, nandi2011guided, liu2009spreadsheet}.
\name{}'s direct manipulation interface enables users to directly work with and modify an existing enriched table to update its associated queries.
For example, imagine a user, Jane, who would like to further explore the result in Figure~\ref{fig:table-screenshot}.
To see the detailed information about the authors of a particular paper, she clicks on its ``author count'' button (Figure~\ref{fig:operators-example}-b).
This simple interaction of tapping the button is translated into a series of \textit{primitive operators} behind the scene,
such as \textit{Select}, as in selecting the row associated with a paper;
and \textit{Add}, as in adding and joining the \texttt{Authors} table with the \texttt{Papers} table.
With a few rounds of similar interactions,
Jane can incrementally build complex queries.

\name{}'s novel ideas work together to address an important, often overlooked problem in databases.
The seminal vision paper by Jagadish et al.~\cite{jagadish2007making} introduced the notion of the presentation data model and argued the importance of direct manipulation interface.
However, designing an easy-to-use system that meets these requirements is challenging. 
\name{} is one of the first instantiations of this important idea, filling a critical research gap, by effectively integrating HCI principles to greatly improve database usability.  
To enable the creation of such a usable tool,
\name{} tightly integrates:
(1) a novel hybrid data model representation, which advances over the relational and nested-relational models, to naturally represent entities and relationships; and
(2) a novel set of interactions that closely work with  the representation to enable users to specify expressive queries through direct manipulation.
With \name{}'s user interface, non-experts can easily and naturally explore databases without writing SQL, while \name{} internally performs queries under the hood.

Through \name{}, we contribute:
\begin{itemize}[noitemsep,topsep=0pt]
\item A novel \textbf{presentation data model} that presents a query result as an enriched table for users to easily browse and explore relational databases 
(Section~\ref{sec:overview}, \ref{sec:model});

\item A \textbf{graph-based model}, called \textit{typed graph model (TGM)} that provides an abstraction of relational databases, for users to explore data in \name{} at a conceptual level (Section~\ref{sec:database});

\item A set of \textbf{user-level actions}, operations that users can directly apply to an enriched table to incrementally construct complex queries and navigate databases (Section
\ref{sec:interactions});

\item The \textbf{usable interface} of \name{} that outperforms a commercial graphical query builder in a \textbf{user study}, in both speed and subjective ratings across a range of database querying tasks  (Section~\ref{sec:interface}, \ref{sec:experiments}).

\end{itemize}

\section{Related Work}
\label{sec:related}

\subsection{Database Usability \& Query Specifications}

Since Query-by-Example (QBE) was developed in 1970s~\cite{zloof1977query}, database researchers have studied fairly extensively the usability aspect of database systems~\cite{jagadish2007making, catarci2000happened, abiteboul2005lowell, idreos2015overview}.
Usability is important, especially because not all database users have expertise in writing complex queries;
many non-technical users find it challenging to write even very simple join queries~\cite{jagadish2007making, abadi2014beckman}.
Many existing approaches are aimed at assisting users with formulating queries.
One representative method is the \textit{visual query builder}, which enables users to visually manipulate schema elements on a graphical interface~\cite{catarci1997visual}.
However, most visual querying systems require that users have precise knowledge of a schema, which makes it difficult for non-experts to use. 
This limitation can be relieved in \textit{keyword search} systems, studied extensively in the last decade~\cite{hristidis2002discover, bhalotia2002keyword, agrawal2002dbxplorer, chen2009keyword},
or natural language interfaces~\cite{li2014constructing}.
However, most of existing approaches~\cite{jayapandian2008automated,fan2011interactive} separate queries and results so that users cannot directly refine query results, which decreases the usability of the systems.
Nandi and Jagadish~\cite{nandi2011guided} argued that users' querying process is often iterative, so database systems should guide users toward interactively formulating and refining queries.

\subsection{Direct Manipulation \& Iterative Querying}

Several database researchers argued that the usability of database querying systems can improve by adopting the \textit{direct manipulation} paradigm~\cite{shneiderman1983direct}, a well-established design principle in the HCI and information visualization areas.
Acknowledging that users' needs are often ambiguous rather than precisely specifiable,  researchers have developed many tools that enable users to interactively browse and explore databases~\cite{idreos2015overview,buoncristiano2015database,singh2016dbexplorer}.
Although they are not specifically designed for relational databases,
a number of interactive visualization systems for entity-relationship data have been developed by information visualization researchers~\cite{kang2007netlens, dunne2012graphtrail, dork2012pivotpaths, liu2011network}. 
For example, NetLens~\cite{kang2007netlens} visualizes relationships between two selected entity types in many-to-many relationships,
and GraphTrail~\cite{dunne2012graphtrail} visually summarizes each entity type and enables users to switch between entities.
Although these visualization systems provide an overview  of data sets, they are not suited for examining database instances along with attributes. 
In exploring and analyzing instance-level information, tabular interfaces, including spreadsheets, are better suited and often preferred by database users~\cite{few2004show,tyszkiewicz2010spreadsheet,liu2009spreadsheet,chang2016using,gonzalez2010google}.
Tyszkiewicz~\cite{tyszkiewicz2010spreadsheet} argued that spreadsheets can play a role as a database engine by using functions and macros.
Liu and Jagadish~\cite{liu2009spreadsheet} formally defined operators that interactively perform grouping operations within a spreadsheet.
However, since the rigid tabular structure does not effectively present many-to-many relationships, the spreadsheet suffers from the same problems  that relational databases have (i.e., a large number of duplications).
To overcome this limitation, Jagadish et al.~\cite{jagadish2011organic} proposed using a presentation view layer on top of underlying databases, which is the notion of the \textit{presentation data model}, defined as a full-fledged layer on top of the logical and physical models.
The challenge is to design presentation data models that help people easily understand join query results and interact with them.

\subsection{Data Models for Effective Presentation}

To develop an intuitive structure for presentation data models, we review a number of data models that conceptualize the mini-world represented in databases.
One such example is the \textit{nested relational model}, studied in the 1980s, which allows each cell  to contain another table that presents one-to-many relationships in a single table~\cite{schek1986relational, roth1988extended}.
The nested model has been used in several studies for designing database interfaces. Bakke et al.~\cite{bakke2016expressive} recently designed a direct manipulation interface for nested-relational databases, and
DataPlay~\cite{abouzied2012dataplay} also used the nested model for presenting query results. 
However, the model suffers from scalability issues because the sizes of the nested tables often become huge when an inner table contains a large number of associated rows or columns~\cite{bakke2013automatic}.
One way to tackle this problem is to replace the inner table with a set of pointers. For example,
the \textit{object-relational model}
lets attributes be user-defined types that include pointers~\cite{stonebraker1995object}. We adapt this idea by introducing an \textit{entity reference} which compactly represents related entities.
Another class of the data models that effectively conceptualize the real-world is the graph data model~\cite{angles2008survey, gyssens1994graph, catarci1993fundamental, sun2012mining}. It represents entities as nodes and relationships as edges based on the \textit{entity-relationship model}~\cite{chen1976entity, batini1992conceptual}.
Catarci et al.,~\cite{catarci1997graphical} used a graph-style \textit{translation layer} for their visual querying system.
To provide users with an easy-to-understand view at an entity-relationship level, we also maintain a graph-style model, transformed from relational databases, under the presentation view.

\section{Introducing ETable}
\label{sec:overview}

\begin{figure}[!tb]
  \begin{center}
  \includegraphics[width=\linewidth,trim={1.05cm 8.2cm 6.2cm 0.75cm},clip]{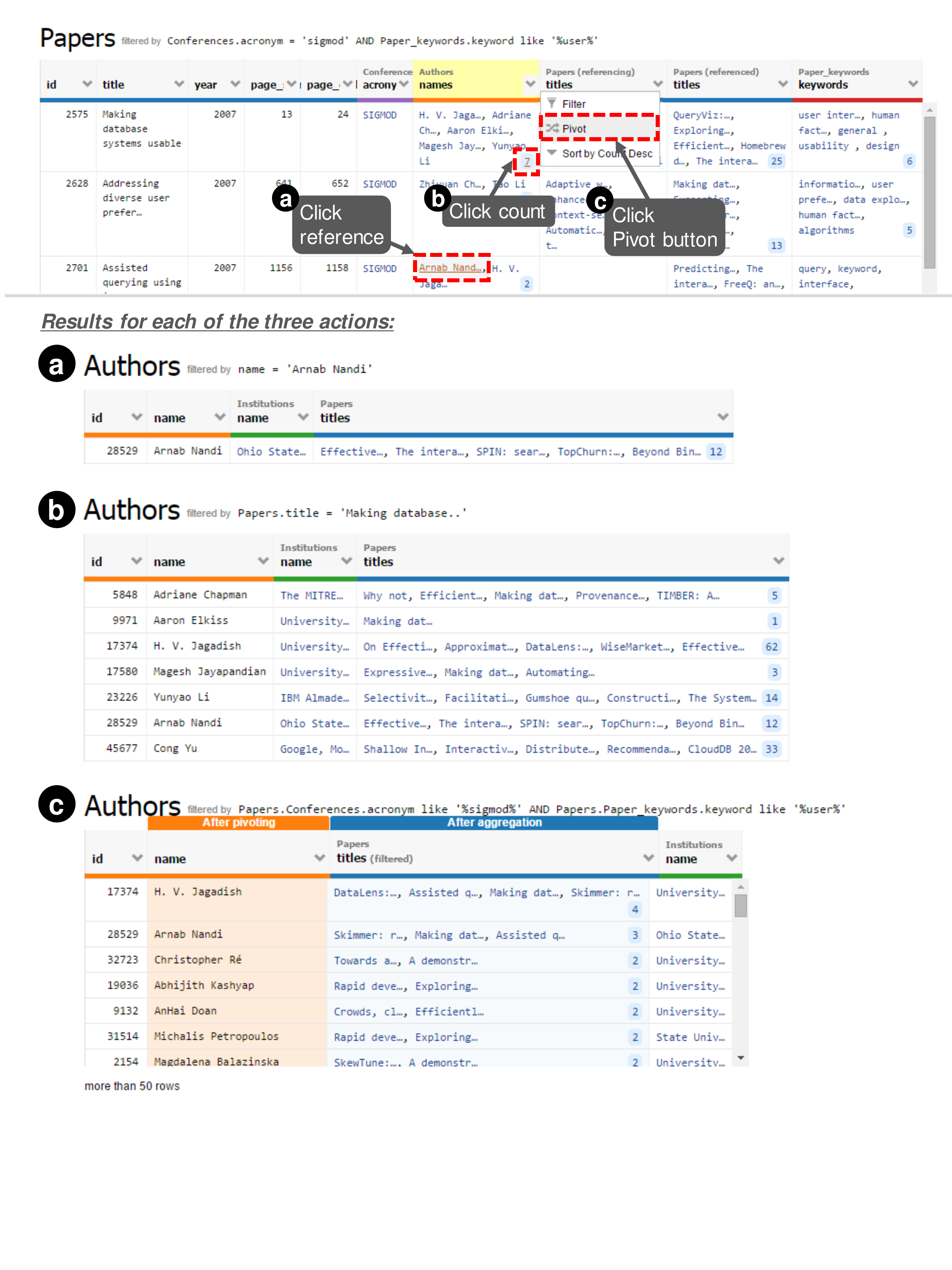}
  \vspace{-17pt}
  \caption{Users can iteratively specify user-level actions by interacting with \name{}. In this example, users can examine further information about paper authors in three ways: (a) clicking on an author's name; (b) clicking a paper's author count; (c) clicking on the pivot button.}
  \label{fig:operators-example}
  \end{center}
  \vspace{-10pt}
\end{figure}

\begin{table*}[t]
\centering
\begin{tabular}{ l  l  l  }
\toprule			
\textbf{Form} & \textbf{Source} & \textbf{Determining factor for mapping from a relational table} \\
\midrule			
Node types & Entity tables & Relation with a single-attribute primary key \\
 & Multi-valued attributes & Relation with two attributes; one of them is a foreign key of an entity relation \\
 & Single-valued categorical attributes & Attribute of low cardinality \\
 \midrule			
Edge types & One-to-many relationships & Foreign key between two entity relations \\
 & Many-to-many relationships & Relation with a composite primary key; both are foreign keys of entity relations \\
 & Multi-valued attributes & From an entity table to a multi-valued attribute \\
 & Single-valued categorical attributes & From an entity table to a categorical attribute \\
\bottomrule  
\end{tabular}
\vspace{-7pt}
\caption{Categories of node and edge types based on how they are translated from relational schema}
\label{table:translate}
\end{table*}

Before we describe the technical details of the proposed data models, we introduce \name{} by describing what users see and how they can interact with it.

\textbf{Representation.}
Figure~\ref{fig:table-screenshot} illustrates an enriched table that we call \textit{Etable}.
As mentioned earlier, it presents a list of SIGMOD papers containing the keyword ``user'' from our collected database (see Figure~\ref{fig:schema} for schema).
Each row of \textit{Etable} represents a single entity of the selected entity type (i.e., \texttt{Papers});
its column represents either a base attribute of the entity (e.g., year) or a set of relevant entities (e.g., authors, keywords).
This representation is formed by pivoting a query result of a join of multiple tables (e.g., \texttt{Papers}, \texttt{Paper\_keywords}, \texttt{Authors}) to a user-selected entity type (e.g., \texttt{Papers}).
One advantage of this representation is that it can simultaneously present all relevant information about an entity in a single row (e.g., authors, keywords, citations).
The relational model cannot represent all of this information in a single relation without duplications because every attribute value must be atomic. 
For instance, when the \texttt{Papers} table is joined with the \texttt{Authors} table, the paper information is repeated as many times as the number of authors, which prevents users from quickly interpreting the results.
We integrate information spread across multiple tables  into a single table by allowing each cell to contain a set of references to other entities.

\textbf{Interactions.}
Users can interact with \textit{Etable} to explore further information.
For instance, to examine further information about the authors of the papers in Figure~\ref{fig:table-screenshot}, 
users can create a new \textit{Etable} that lists authors in several ways, as depicted in Figure~\ref{fig:operators-example}:
(1) If users are interested in one of the authors (e.g., Arnab Nandi), they can click on his name to create a new \textit{Etable} consisting of one row that presents its attributes; (2) if users want to list the complete set of authors (e.g., all seven authors of the paper titled ``Making database systems usable''), they can click on the author count in the right corner of the cell (i.e., 7); and (3) if users want to list and sort the entities across the entire rows in a column (e.g., Who wrote the most papers about ``user'' in SIGMOD?), they can click on the \textit{pivot} button on the column menu, which groups and sorts the authors based on the number of papers they have written.
By gradually applying these operations, users can incrementally make sense of data and build complex queries.

\section{Typed Graph Model}
\label{sec:database}

In this section, we define a \textit{typed graph model (TGM)} which enables users to explore relational databases on a conceptual entity-relationship level without having to know a logical schema. 
A relational schema and instances are translated into a \textit{database schema graph} and \textit{database instance graph} as a preprocessing step, and
all operations specified by users on the \name{} interface are executed over these graphs, not relational databases.

We represent entities and relationships as a graph with types and attributes. Each entity forms a node, and relationships among the entities become edges. A \textit{typed graph database (TGDB)} consists of a \textit{TGDB schema graph}, $\mathcal{G}_S$, and a \textit{TGDB instance graph}, $\mathcal{G}_I$.

\begin{figure}[!bt]
  \begin{center}
  \includegraphics[width=\columnwidth,trim={1.0cm 1.6cm 0.2cm 5.3cm},clip]{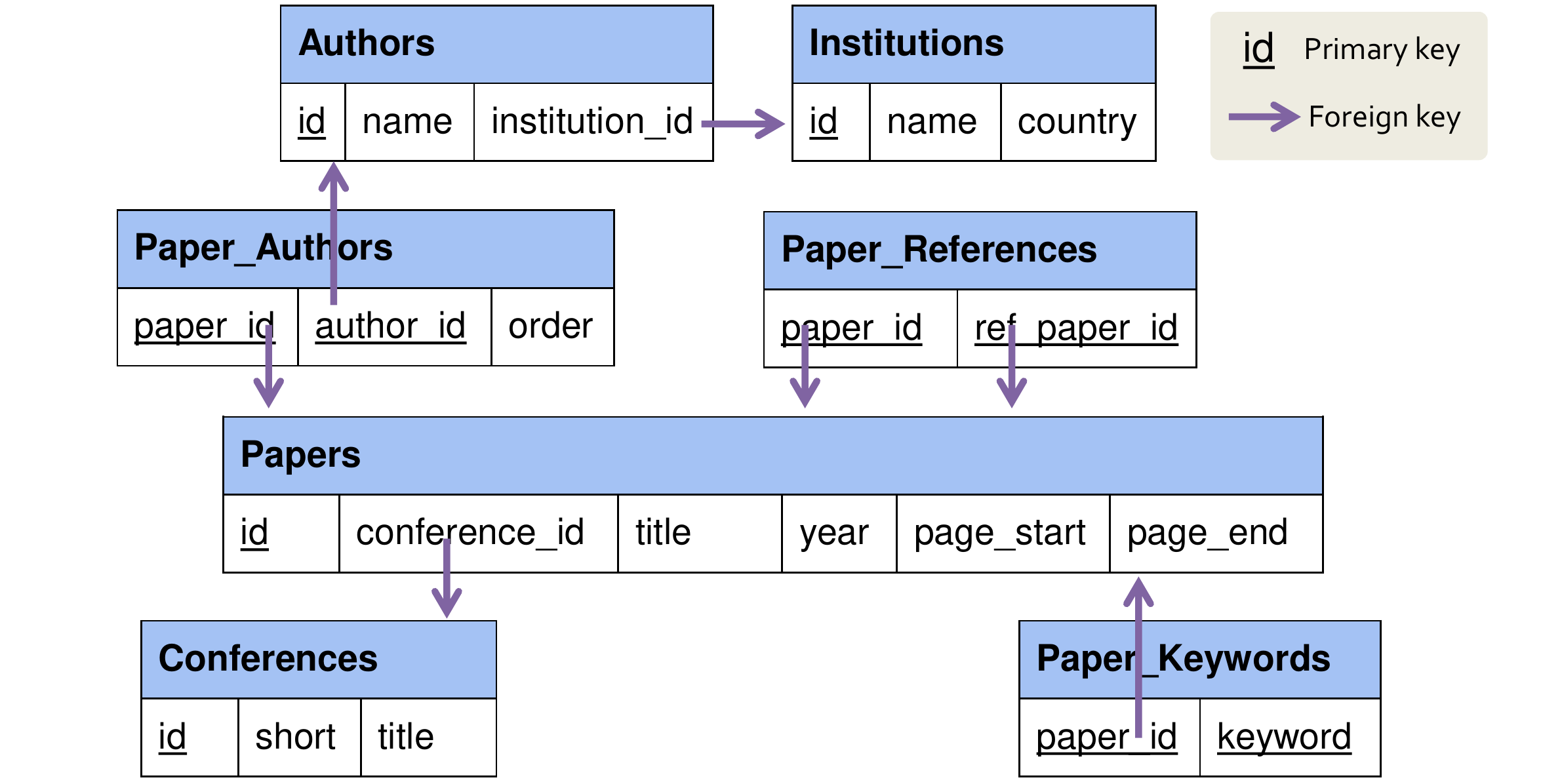}
  \vspace{-17pt}
  \caption{The relational schema of the academic data set used in this work, 7 relations in total. 
  }
  \label{fig:schema}
  \end{center} 
  \vspace{-12pt}
\end{figure}

\begin{definition}
\textbf{Schema Graph.}
A TGDB schema graph $\mathcal{G}_S$ is a tuple $(\mathcal{T}, \mathcal{P})$,
where $\mathcal{T}$ represents a set of node types (or entity types\footnote{We use the words ``node'' and ``entity'' interchangeably. A node is used more formally; an entity is used more for presentation to users.}), and $\mathcal{P} \subseteq \mathcal{T} \times \mathcal{T}$ represents a set of edge types (or relationship types).
Each node type $\tau_i \in \mathcal{T}$ is a tuple $(\alpha_i, \mathcal{A}_{i}, \beta_{i})$, where $\alpha_i$ denotes the name of a node type, $\mathcal{A}_{i}$ is a set of single-valued attributes, and $\beta_{i}$ is a label attribute chosen from one of the attributes and used to represent node instances of this type.
Each edge type $\rho \in \mathcal{P}$ also has a name and a set of attributes. 
We denote the source and target node types of $\rho$ as $source(\rho)$ and $target(\rho)$, respectively.
All the edge types, except self loops, are bidirectional. 
\end{definition}

\begin{definition}
\textbf{Instance Graph.}
A TGDB instance graph $\mathcal{G}_I$, is a tuple $(V, E)$, where $V$ represents a set of nodes (or entities) and $E$ represents a set of edges (or relationships) between two nodes.
Every instance graph $\mathcal{G}_I$ has a corresponding schema graph $\mathcal{G}_S$, and the instance graph has a node type mapping function $type_{\tau}: V \rightarrow \mathcal{T}$ and an edge type mapping function $type_{\rho}: E \rightarrow \mathcal{P}$
that partition nodes $V$ into $V_1, ..., V_{n_\mathcal{T}}$ and edges $E$ into $E_1, ..., E_{n_\mathcal{P}}$.
Each node $v_{} \in V$ consists of a set of attribute values $v[A_{ij}]$ for the attributes of the corresponding node type and has a label defined as $label(v) = v[\beta_i]$. 
Each edge $e \in E$ consists of a set of attribute values $e[A_{ij}]$ for its type.
We denote the source and target nodes of $e$ as $source(e)$ and $target(e)$, respectively.
\end{definition}

\begin{figure}[!bt]
  \begin{center}
  \includegraphics[width=\columnwidth,trim={1.2cm 3.2cm 0cm 8.3cm},clip]{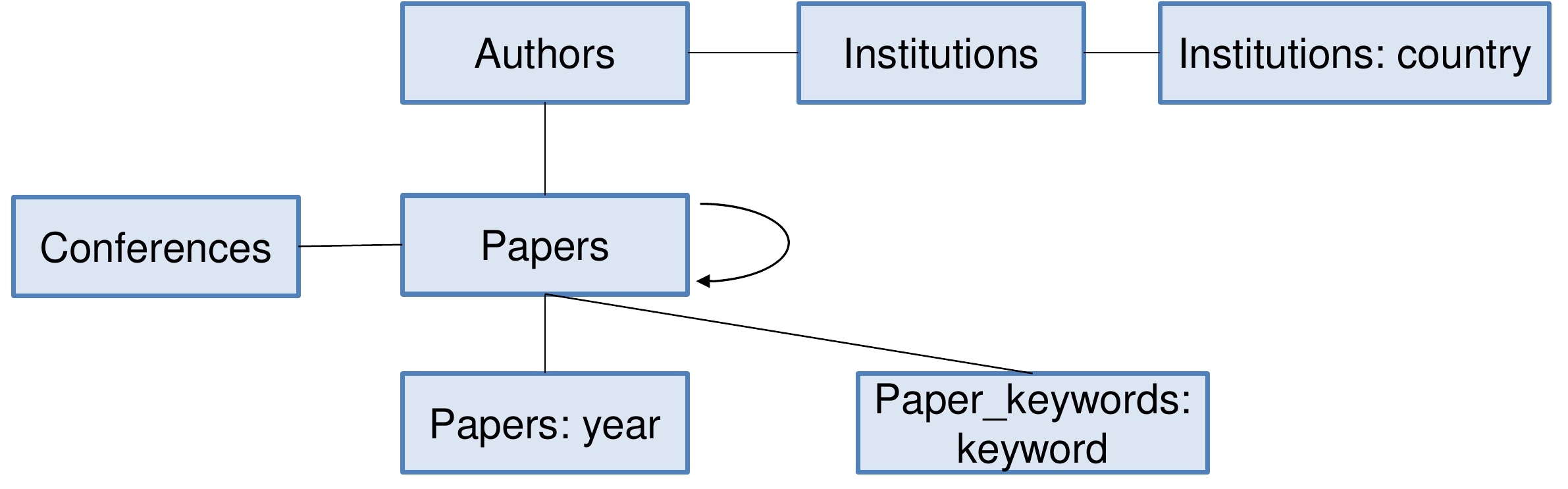}
  \vspace{-17pt}
  \caption{TGDB schema graph constructed from the relational schema in Figure~\ref{fig:schema}. 
  Each rectangle represents a node type, 
  and each edge is an edge type.
  }
  \label{fig:schema-graph}
  \end{center} 
  \vspace{0pt}
\end{figure}

\begin{figure}[!bt]
  \begin{center}
  \includegraphics[width=\columnwidth,trim={2.1cm 2.8cm 2.5cm 3.2cm},clip]{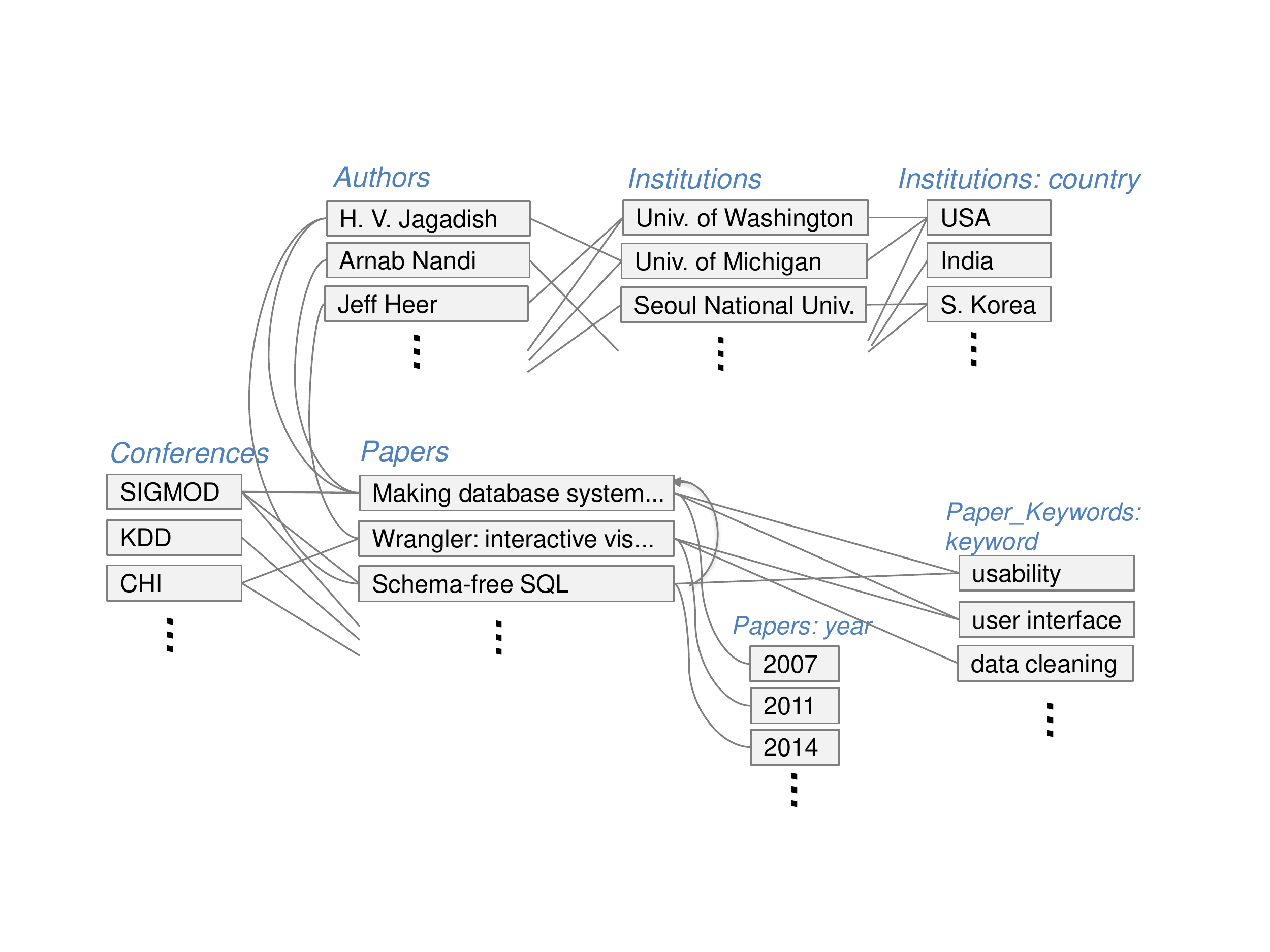}
  \vspace{-19pt}
  \caption{
  A part of the TGDB instance graph constructed from the academic data set used in this paper, following the schema in Figure~\ref{fig:schema-graph}. 
  Node types shown in blue italic font.
  }
  \label{fig:instance-graph}
  \end{center} 
  \vspace{-7pt}
\end{figure}

The typed graph model, similar to many graph data models~\cite{angles2008survey, gyssens1994graph, sun2012mining}, is much more effective for conveying a  conceptual understanding of the mini-world represented in databases than the relational model.   
As it abstracts relational databases, users can ignore the logical and physical representation of data.
Users can also easily understand the structure of data,
since nodes always represent entities and edges represent relationships, 
Unlike TGM, the relational model is a mixture of entities, relationships, and multivalued attributes.
Although some existing graph models are more expressive for representing a variety of relationships (e.g., hierarchical parent-child relationships among entities), we simply use nodes and edges to focus on making the semantics of the underlying relations more explicit by mapping to entities and relationships that they represent in the real world.

Relational databases can be translated into the TGDB schema and instance graphs in a near-automatic process. 
We adapt the reverse engineering literature pertaining to translating relational data\-bases into several graph-style models~\cite{batini1992conceptual, chiang1994reverse, sequeda2012directly}.
A detailed procedure presented in Appendix \ref{sec:appendix} includes an analysis of a relational schema based on primary keys, foreign keys, and cardinalities for classifying tables into several categories, and a series of actions that create the schema graph.
Table~\ref{table:translate} summarizes the categories of node and edge types based on how they are determined from relational schema. 
Figures~\ref{fig:schema-graph} and \ref{fig:instance-graph} illustrate a schema graph and a part of the instance graph constructed from an academic publication database whose schema is shown in Figure~\ref{fig:schema}.

\section{ETable Presentation Data Model}
\label{sec:model}

We present our \name{} presentation data model for usable exploration of entities and relationships in databases.

\subsection{Enriched Table}

A query result in the \name{} model is presented as an enriched table, which we also call \name{}.
An \name{} $R$ has a set of columns $\mathcal{A}$ and consists of a set of rows $r \in R$.
The columns are categorized into two types: \textit{single-attribute columns} and \textit{entity-reference columns}.
The value of the single-attribute column $r[A]$ is atomic as it is in the relational model.
The value of the entity-reference column $r[A]$ contains a single or a set of \textit{entity references}. 
The entity reference refers to another node in the database instance graph.
Unlike a foreign key in the relational model,
each entity reference is shown as a clickable label, similar to a \textit{hyperlink} on a webpage.
Just like how a hyperlink's hypertext describes the webpage that the link points to (instead of its URL),
for example, \name{} represents an author's entity reference by the author name (instead of the author ID).

The entity-reference columns present rich information spread across multiple relations within a single enriched table.
While a foreign key attribute in the relational model  contains only a single reference for a many-to-one relationship because of the first normal form,
an entity-reference column can represent one-to-many relationships, many-to-many relationships, or multivalued attributes in a single column.
Furthermore, the entity-reference column has advantages over the nested relational model which
requires much screen space as it squeezes another table into cells, leading to inefficient browsing.
Unlike the nested model, \name{} presents clickable labels that compactly show information and allow users to further explore relevant information.

\subsection{ETable Specification}

An \name{} can be specified by selecting specific elements of the TGDB database schema and instance graphs introduced in the previous section.

\begin{definition}
\textbf{ETable Query Specification.}
An \name{} $R$ is specified by a \textit{query pattern} $Q$, which is a tuple 
$(\tau_a, T, P, \mathcal{C})$.

\begin{enumerate}
\item \textbf{Primary node type} $\tau_a$: It is one of the node types in the schema graph.
Each row of \name{} will represent a single node instance of the primary node type.
\item \textbf{Participating node types} $T$:
It is a set of node types chosen from the node types in the schema graph (i.e., $T = \{t_1, ..., t_{n_T}\}, \forall t_i \in \mathcal{T}$).
It must contain the primary node type $\tau_a$ (i.e., $\tau_a \in T$).
It determines the scope of data instances and is similar to a set of relations in SQL FROM clauses.
A node type in the schema graph can exist multiple times in the participating node types, like a relational algebra expression can contain the same relation multiple times.
\item \textbf{Participating edge types} $P$:
It is a set of edge types selected from the schema graph (i.e., $P = \{p_1, ..., p_{n_P}\}, \forall p_i \in \mathcal{P}$). 
It connects the participating nodes types, thus
all the source and target nodes of these edges should exist in the participating node types
(i.e., $source(p_i) \in T \wedge target(p_i) \in T,  \forall p_i \in \mathcal{P}$). 
\item \textbf{Selection conditions for node types} $\mathcal{C}$: 
It is a set of selection conditions $\mathcal{C} = (C_1, ..., C_{n_T})$ applied to each of the participating node types, i.e., $C_i$ applies to $t_i \in T$. 
\end{enumerate}
\end{definition}

\begin{figure}[!t]
  \begin{center}
  \includegraphics[width=1.0\linewidth,trim={0.3cm 15.9cm 15.0cm 0.9cm},clip]{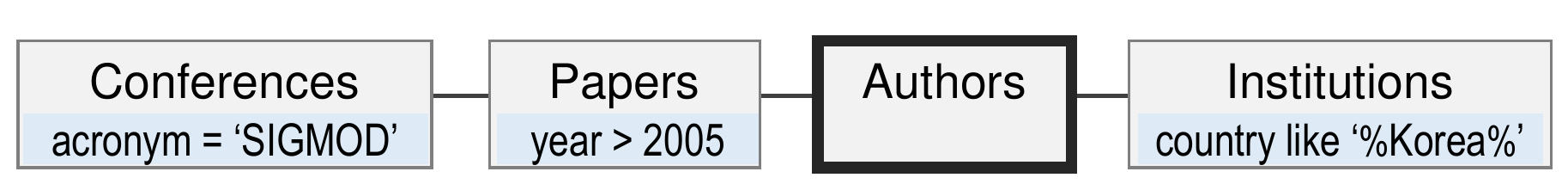}
  \vspace{-14pt}
  \caption{An example query pattern in a diagrammatic notation. It represents a query that finds a list of researchers who have published papers at SIGMOD after 2005 and are currently working at institutions in Korea.}
  \label{fig:query-pattern}
  \end{center}
  \vspace{-2pt}
\end{figure}

A query pattern can be represented as an acyclic graph where one of the nodes is marked as a primary node type and any node can have selection conditions. 
For example, the query pattern in Figure~\ref{fig:query-pattern} represents a query that produces a list of researchers  who have published papers at SIGMOD after 2005 and are currently working at institutions in Korea.

\begin{figure*}[!t]
  \begin{center}
  \includegraphics[width=1.0\linewidth,trim={6.8cm 15.0cm 16.9cm 0.7cm},clip]{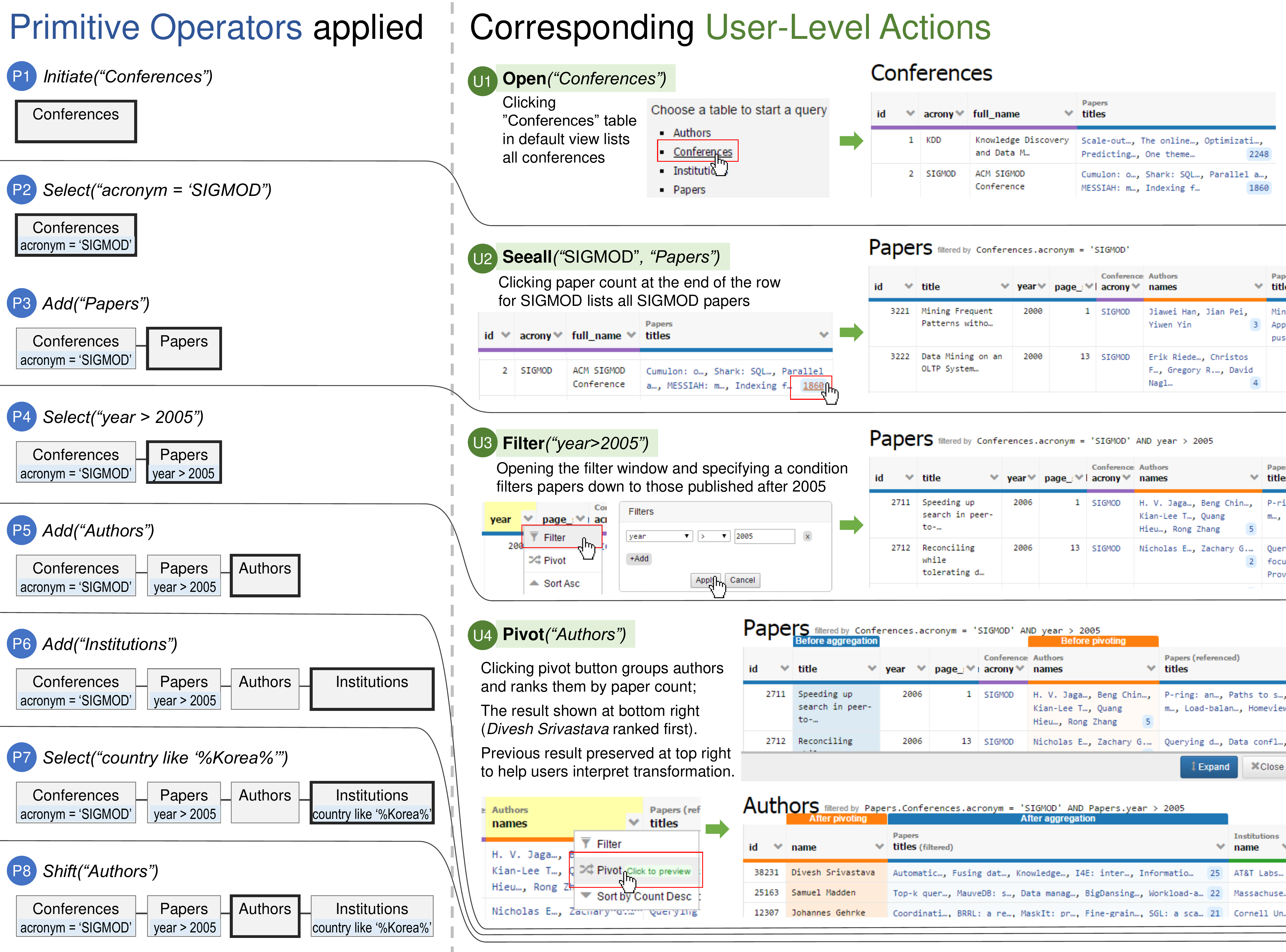}
  \vspace{-15pt}
  \caption{An example of incrementally building a complex query: \textit{find a list of researchers who have published papers at SIGMOD after 2005 and are currently working at institutions in Korea}. 
  \textbf{Left}: constructing the query through a series of \name{} primitive operators. 
  \textbf{Right}: corresponding user actions in the interface that invoke the operators (Section~\ref{sec:interactions} describes the user-level actions in detail). 
  User actions for the operators P6-P8, similar to the others shown in the figure, are omitted for brevity.}
  \label{fig:session-example-both}
  \end{center}
  \vspace{-2pt}
\end{figure*}

\begin{figure*}
  \begin{center}
  \includegraphics[width=1.0\linewidth,trim={0.3cm 6.7cm 0.3cm 4.4cm},clip]{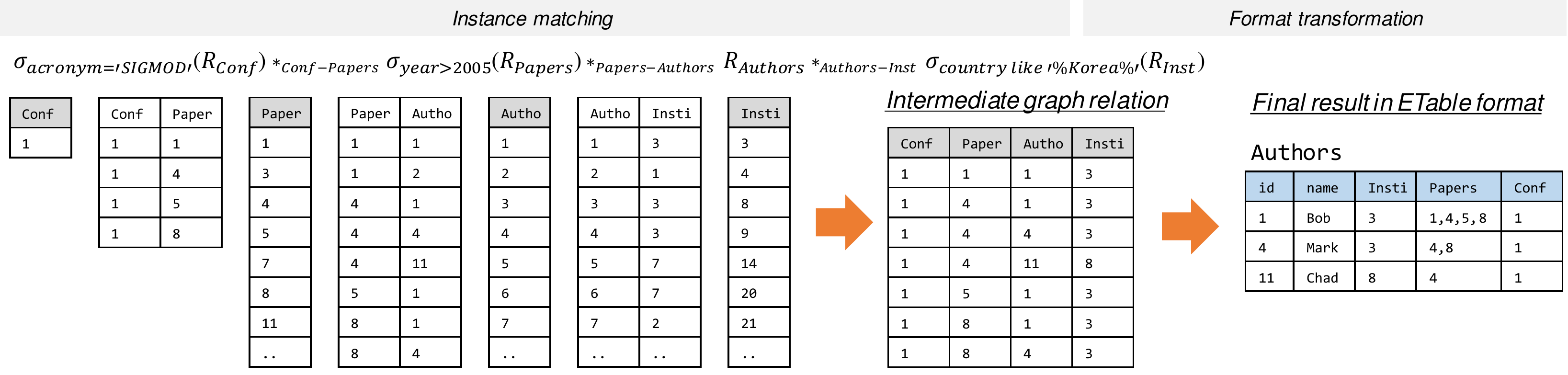}
  \vspace{-17pt}
  \caption{\name{} query execution process consists of two steps: (1) the instance matching step which extracts matched instances from the instance graph and (2) the format transformation step which transforms the instances into the \name{} format.}
  \label{fig:execution-example}
  \end{center}
  \vspace{-7pt}
\end{figure*}

\subsection{Incremental Query Building with Primitive Operators}
\label{sec:operations}

In \name{}, a query pattern can be constructed and updated by \textit{primitive operators}.
Each operator builds on an existing \name{} query to generate a new, updated \name{} query.
In this subsection, we describe these operators in detail.
In Section~\ref{sec:interactions}, we will describe how users' actions performed on the \name{} user interface will invoke these operators.
Formally, given an \name{} specification $Q(\tau_a, T, P, \mathcal{C})$, each of the following operator creates a new specification $Q'(\tau'_a, T', P', \mathcal{C}')$, except the \textit{Initiate} operator which creates a new \name{} from scratch.

\begin{enumerate}

\item \textbf{Initiation.}
A new \name{} can be created by selecting one of the node types $\tau_k$ in the schema graph. Its result lists the corresponding nodes. 
\begin{align*}
&Initiate(\tau_k) = Q' \\
&\text{where }  
\tau'_a = \tau_k ,\text{ }
T' = \{\tau_k\} ,\text{ }
P' = \{\} ,\text{ }
\text{and } \mathcal{C} = \{\}.
\end{align*}

\item \textbf{Selection.}
\name{} rows can be filtered based on their columns, similar to the selection operator in the relational model.
Applying a selection condition $C_k$ to the primary node type $\tau_a$ filters the rows of the current \name{}.
\begin{align*}
&Select(C_k, Q) = Q' \\
&\text{where }  
\tau'_a = \tau_a ,\text{ }
T' = T ,\text{ }
P' = P ,\text{ }
\text{and } C'_a = C_k.
\end{align*}

\item \textbf{Adding a node type.}
Another node type can be added to a query pattern to examine how it is related to the current primary node type.
It corresponds to adding a join operator in the relational model.
Selecting one of the node types that are linked to the primary node type $\tau_a$ by an edge type $\rho_k$ (i.e., $source(\rho_k) = \tau_a$), adds it to the participating node types in the current query $Q$.
\begin{align*}
&Add(\rho_k, Q) = Q'\\
&\text{where }  
\tau'_a = target(\rho_k) ,\text{ }
T' = T \cup \{target(\rho_k)\} ,\text{ }\\
&\quad\quad\quad P' = P \cup \{\rho_k\} ,\text{ }
\text{and } \mathcal{C}' = \mathcal{C} \cup \{\}.
\end{align*}

\item \textbf{Shifting focus to another participating node type.}
The primary node type $\tau_a$ can be changed to one of the other participating node types $\tau_k$. 
It can be thought of as representing the current join result from a different angle.
\begin{align*}
&Shift(\tau_k, Q) = Q'\\
&\text{where } 
\tau'_a = \tau_k ,\text{ }
T' = T ,\text{ }
P' = P ,\text{ }
\text{and } \mathcal{C}' = \mathcal{C}.
\end{align*}

\end{enumerate}

The above primitive operators enable us to build any complex queries by incrementally specifying the operators one-by-one.
Figure~\ref{fig:session-example-both} (left) illustrates the query construction process consisting of 8 operators.
A new query pattern can be created with \textit{Initiate}; 
Selection conditions can be added with \textit{Select}, just like writing expressions in WHERE clauses in SQL;
and node types can be added with \textit{Add}, just like adding relations to FROM clauses and setting one of them as a GROUP BY attribute.
Also, the primary node type can be changed with \textit{Shift}, similar to changing the GROUP BY attribute.
A sequence of these operators specified constitutes a query pattern in the \name{} model. 
These operators can be invoked by users on the user interface with \textit{user-level actions}, 
which we will describe details in Section~\ref{sec:interactions}. 
The right side of Figure~\ref{fig:session-example-both} shows how users can specify the same query through the user interface.

\subsection{Query Execution}
\label{sec:execution}

A query pattern is executed to produce a result in the \name{} format. The execution process is divided into two steps: \textit{instance matching} and \textit{format transformation}. The first step extracts matched node instances from the TGDB instance graph, and the second step transforms a result from the first step into the \name{} format. 

\subsubsection{Instance Matching}

The instance matching process finds a set of matched instances for a given query pattern.
Formally, it returns a \textit{graph relation} $R^G$, which consists of a set of tuples, each of which contains a list of node instances in the database instance graph.
The graph relation is generated with an \textit{instance matching function} $m(Q)$, which consists of a series of operations.
The operations constitute primitives which make up a \textit{graph relation algebra}.

A \textit{graph relation} $R^G$, similar to a relation in the relational model, consists of a set of tuples with a set of attributes. 
The schema of the graph relation is defined as a set of node types $\mathcal{A} = (A_1, ..., A_n)$ where $A_i \in \mathcal{T}$.
In other words, each attribute $A_i$ corresponds to a node type.
The node type $\tau_j$ determines the domain of the attribute (i.e., ${domain}_i = \{ v | v \in V_j \}$).
A \textit{base graph relation} is defined as a graph relation with a single attribute. In other words, each node type $\tau_1, ..., \tau_n$ produces a base graph relation $R^G_1, ..., R^G_n$. 
A \textit{non-base graph relation} can be created by applying the following graph relation operators to the base graph relations.

\begin{enumerate}
\item \textbf{Selection.}
It filters tuples of a graph relation $R$ using a selection condition $C_i$ applicable to one of the node types $A_i$.
\begin{equation*}
\sigma_{C_i}(R^G) = \{ r | r \in R^G
 \wedge r[A_i] \text{ satisfies } C_i \}.
\end{equation*}

\item \textbf{Join.}
It joins two graph relations $R_1$ and $R_2$ using edge types $\rho_k$. The attributes of the created graph relation is a concatenation of the attributes of the two graph relations.
\begin{align*}
R^G_1 \ast_{\rho_k} R^G_2 &= \{ (r_1, r_2) | r_1 \in R^G_1 \wedge r_2 \in R^G_2 \\ & \wedge 
source(\rho_k) \in \mathcal{A}_1 \wedge target(\rho_k) \in \mathcal{A}_2   \}  .
\end{align*}
We use a symbol, $\ast$, to differentiate it from the relational correspondence, $\bowtie$, and not to be confused with natural join.

\item \textbf{Projection.}
It removes all attributes of the graph relations except the given attribute. Duplicated rows are eliminated.
\begin{equation*}
\Pi_{A_i}(R^G) = \{ r[A_i] | r \in R^G \}.
\end{equation*}

\end{enumerate}

These operators enable us to define an instance matching function $m(Q)$. In fact, this function only requires the \textit{Selection} and \textit{Join} operators: the \textit{Projection} operator will be used later in the format transformation step.

\begin{definition}
\textbf{Instance Matching.}
Given a \name{} query pattern $Q(\tau_a, T, P, \mathcal{C})$, a matching function $m$ returns a graph relation $R^G$ containing node instances in the instance graph  $\mathcal{G}_I$.
\begin{equation*}
m(Q) = \sigma_{C_1} (R^G_1) \ast_{p_1} \sigma_{C_2} (R^G_2) \ast_{p_2} ... \ast_{p_{n-1}} \sigma_{C_n} (R^G_n) ,
\end{equation*}
where $R^G_i$ is a base graph relation obtained from a node type $t_i \in T$, i.e., $R^G_i = \{ v |v \in V \wedge type(v) = t_i \}$,
$C_i \in \mathcal{C}$ is a selection condition for $R_i$, and $p_i \in P$ is one of the edge types that joins graph relations on both sides, i.e., $p_i = \{ p | p \in P \wedge source(p) \in \{ t_1, ... t_i \} \wedge target(p) \in \{ t_{i+1}, ... t_n \} \}$. 

\end{definition}

Figure~\ref{fig:execution-example} (left) illustrates the instance matching process. It returns a graph relation, which is an intermediate format to be transformed into the \name{} format.

\begin{figure*}
  \centering
  \includegraphics[width=\linewidth,trim={0cm 6.7cm 2.3cm 0cm},clip]{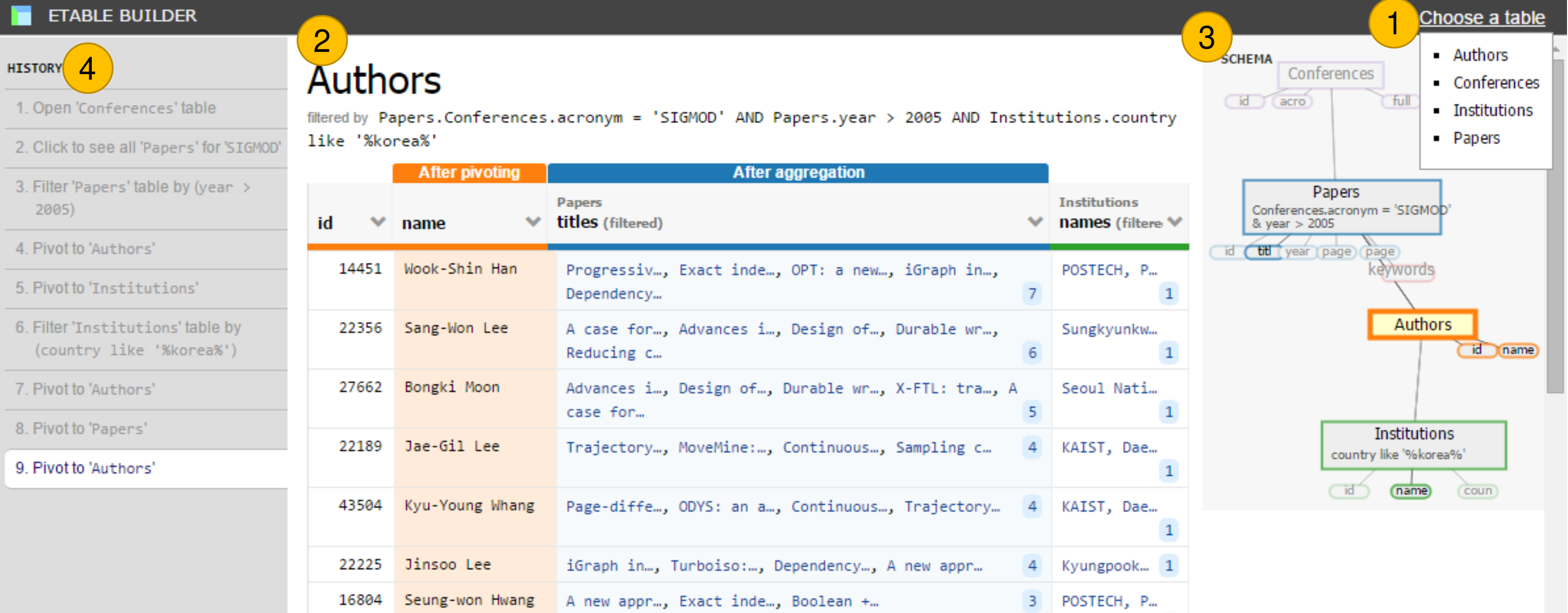}
  \vspace{-16pt}
  \caption{The \name{} interface consists of (1) the default table list for initiating a query, (2) the main view presenting query results, (3) the schema view showing a query pattern, and (4) the history view listing  operators specified by users. Users can build queries and explore databases by directly interacting with the interface.}
  \label{fig:screenshot}
\end{figure*}

\subsubsection{Format Transformation}
\label{sec:execute-2}

A graph relation obtained from the \textit{instance matching} function is transformed into the \name{} format.
We describe how rows and columns of \name{} are determined from it. 

The rows of \name{} consist of nodes of the primary node type, filtered by all selection conditions in the query pattern.
They are extracted from the instance matching result:
\begin{equation*}
R = \{ v | v \in \Pi_{\tau_a} ( m(Q(\tau_a, T, P, \mathcal{C}))) \} .
\end{equation*}
Given the result of the instance matching function, all attributes except the attribute representing the primary node type are discarded, and then, each of distinct node in that column becomes a row. 

\name{} has three types of columns to present rich information for each row. 
In addition to the attributes of the primary node types, which we call \textit{base attributes} $\mathcal{A}_b$, we introduce two other types of columns for presenting a set of entity references: \textit{participating node columns}, $\mathcal{A}_t$, and \textit{neighbor node columns}, $\mathcal{A}_h$.

\begin{enumerate}
\item \textbf{List of base attributes} $\mathcal{A}_b$: It is a full set of the attributes $A$ of the primary node type $\tau_a$. 
The value of the column $A_j \in \mathcal{A}_b$ would be a single value:
\begin{equation*}
r[A_j] = v[A_j].
\end{equation*}

\item \textbf{List of participating node types} $\mathcal{A}_t$: It is a set of all the node types $T$ in the query pattern, except the primary node type $\tau_a$,
i.e., $\mathcal{A}_t = \{ \tau | \tau \in T \wedge \tau \neq \tau_a  \}$.
The value of the column $A_j \in \mathcal{A}_t$ would be a set of entity references:
\begin{align*}
r[A_j] =& \{ u | u \in V \wedge A_j = type(u) \\ 
& \wedge \Pi_{type(u)} \sigma_{\tau_a = r} (m(Q)) \}.
\end{align*}

\item \textbf{List of neighbor node types} $\mathcal{A}_h$: It is a set of all the neighboring node types of the primary node type $\tau_a$ in the schema graph regardless of the query pattern, 
i.e., $\mathcal{A}_h = \{ (\rho, \tau) | \tau \in \mathcal{T} \wedge \rho \in  \mathcal{P} \wedge source(\rho) = \tau_a  \wedge target(\rho) = \tau  \}$.
The value of the column $A_j \in \mathcal{A}_h$ would be a set of nodes references:
\begin{align*}
r[A_j] =& \{ u | u \in V \wedge e \in E \wedge A_j = (type(e), type(u)) \\ 
& \wedge  u = target(e) \wedge r = source(e) \}.
\end{align*}

\end{enumerate}
Figure~\ref{fig:execution-example} (right) illustrates the results produced from the format transformation process. The first two columns are base attributes, and the rest of the columns are participating node columns. We omit neighbor node columns as some of these columns are the same as the participating node columns.

By transforming the graph relation into the \name{} format, we compactly present join query results without duplications. 
Each row of \name{} is uniquely determined by a node of a primary node type.
The participating node columns show all the other entity types in the query pattern with respect to the primary node type. 
This transformation process is similar to setting one of the relations as a GROUP BY attribute in SQL, but while GROUP BY aggregates the corresponding instances  into a single value (i.e., COUNT, AVG), \name{} presents a list of the corresponding instances as entity references. 
The neighbor node columns are also useful for describing the rows of the \name{}, although information in these columns is not obtained from the graph relation. These columns enable users to browse one-to-many or many-to-many relationships. Moreover, they provide users with a preview of possible new join operations as it presents all the join candidates.
For instance, a \name{} in Figure~\ref{fig:table-screenshot} consists of many neighbor node columns (e.g., Authors) that helps users browse rich information about each paper.

\section{Interface \& System Design}
\label{sec:interface}

\name{}'s  interface (Figure~\ref{fig:screenshot}) consists of four components: (1) the default table list, (2) the main view, (3) the schema view, and (4) the history view.
The \textit{default table list} presents a list of entity types in the schema graph. Users can pick one from the list to initiate a query. 
The \textit{main view} presents an \name{} executed based on a query pattern which is graphically shown over the \textit{schema view}. Users can directly interact with the main view to update the current query. The list of actions specified by users is presented on the \textit{history view}, which allows users to revert to a previous state.

\subsection{User-Level Actions}
\label{sec:interactions}

\renewcommand{\arraystretch}{1.1}
\begin{table*}[t]
\centering
\begin{tabular}{ l  c  c }
\toprule			
\textbf{Task} & \textbf{Category} & \textbf{\#Relations}\\
\midrule			
1. Find the year that the paper titled `Making database systems usable' was published in. & Attribute & 1 \\
2. Find all the keywords of the paper titled `Collaborative filtering with temporal dynamics'. & Attribute & 2 \\
3. Find all the papers that were written by `Samuel Madden' and published in 2013 or after.  & Filter & 3 \\
4. \parbox[t]{13cm}{Find all the papers written by researchers at `Carnegie Mellon University' and published \\at the KDD conference.} & Filter & 5\vspace{3pt}\\
5. Which institution in South Korea has the largest number of researchers? & Aggregate & 2 \\
6. Find the top 3 researchers who have published the most papers in the SIGMOD conference. & Aggregate & 4 \\
\bottomrule  
\end{tabular}
\vspace{-7pt}
\caption{List of tasks. 
	Task 1 \& 2 retrieve attribute values,
	task 3 \& 4 filter entities,
	task 5 \& 6 perform aggregations. 
}
\label{table:tasks}
\end{table*}

Users can update the current query pattern by directly interacting with \name{} via \textbf{user-level actions}. 
As shown in
Figure~\ref{fig:session-example-both}, these actions in turn invoke their corresponding primitive operators (discussed in Section~\ref{sec:operations}).

\begin{enumerate}
\item \textbf{Open a new table.}
Users can open a new table by clicking a node type $\tau_k$ on the default table list.
The action invokes the $Initiate(\tau_k)$ operator (Fig~\ref{fig:session-example-both}: action U1).
\begin{align*}
&\textbf{Open}(\tau_k) = Initiate(\tau_k).
\end{align*}

\item \textbf{Filter.}
Users can filter the rows of the current \name{} by inducing selection conditions
via a popup window at the column header 
(Fig~\ref{fig:session-example-both}: action U3).
Besides the base attributes, users can also filter rows by the labels of the neighbor nodes columns (e.g., authors' names), which is translated into subqueries.
We currently provide only a conjunction of predicates,
but it is straightforward to provide disjunctions and more operations.
The action invokes the \textit{Select} operator.
\begin{align*}
&\textbf{Filter}(C, R) = Select(C, R).
\end{align*}

\item \textbf{Pivot.}
Users can change the primary node type by clicking the pivot button on the context menu for neighbor or participating node columns.
It calls the \textit{Add} operator if the column is the neighbor node type (Fig~\ref{fig:session-example-both}: action U4); it performs the \textit{Shift} operator if it is the participating node type.
\begin{align*}
&\textbf{Pivot}(\rho_l, R) = Add(\rho_l, R),\\ 
\text{or } &\textbf{Pivot}(\tau_k, R) = Shift(\tau_k, R).
\end{align*}

\item \textbf{See a particular node.}
When users are interested in one of the entity references, they can click it to create a new \name{} consisting of a single row presenting the clicked entity.
Unlike the above actions, it invokes two primitive operators:
it initiates a new \name{}, and then perform the \textit{Select} operator to show the single node. For the clicked node $v_k$: 
\begin{align*}
\textbf{Single}(v_k, R) = &Select(C, type(v_k), Initiate(type(v_k)),\\ 
&\text{where } C = \{u | u = v_k \}. 
\end{align*}

\item \textbf{See all related nodes.}
When users are interested in a full list of entity references, they can click a number (i.e., entity reference count) in the right corner of a cell (Fig~\ref{fig:session-example-both}: action U2).
It also encapsulates two primitive operators.
The operators invoked are different depending on whether the selected column is \textit{neighbor} or \textit{participating} node column.
For the \textit{neighboring node column} $\rho_l$ of $v_k$:
\begin{align*}
\textbf{Seeall}_h(v_k, \rho_l, R) = &Add(\rho_l, Select(C, type(v_k), R)),\\
&\text{where } C = \{u | u = v_k \}, 
\end{align*}
and for the \textit{participating node column} $t_l$:
\begin{align*}
\textbf{Seeall}_t(v_k, t_l, R) = &Shift(t_l, Select(C, type(v_k), R)), R)),\\
&\text{where } C = \{u | u = v_k \} \}. 
\end{align*}

\end{enumerate}
\name{} supports additional actions that help with database exploration, such as:
(1) \textbf{Sort rows} based on the values in a column; 
(2) \textbf{Hide/show columns} to reduce visual complexity in the interface; and 
(3) \textbf{Revert to previous queries} via 
the left history panel.

\subsection{Architecture}
\name{} system uses a three-tier architecture, consisting of 
(1) an interactive user interface front-end that can run in any modern web browsers, written in HTML, JavaScript, and D3.js\footnote{\url{https://d3js.org/}};
(2) a Python-based application server; and 
(3) a PostgreSQL database backend.
The PostgreSQL database stores TGDB schema and instance graphs in four relational tables: \texttt{nodes}, \texttt{edges}, \texttt{node\_types,} and \texttt{edge\_types}.
A query pattern for \name{} is translated into SQL queries that operate on the PostgreSQL database.
To efficiently perform queries, we partition a long SQL query into multiple queries consisting of a fewer number of relations to be joined (i.e., each for a single entity-reference column) and merge them.

\section{Evaluation: User Study}
\label{sec:experiments}

To evaluate the usability of \name{}, we conducted a user study that tests whether users can construct queries quickly and accurately.
We compared \name{} with Navicat Query Builder.\footnote{\url{http://www.navicat.com/}}
Navicat is one of the most popular commercial database administration tools with a graphical query building feature.
Graphical builders such as Navicat Query Builder have been commonly used as baseline systems in database usability research~\cite{liu2009spreadsheet,mandel2013gestural,bakke2016expressive}.

\subsection{Experimental Design}

\textbf{Participants.}
We recruited 12 participants from our university through advertisements posted to mailing lists at our institution. 
All were graduate students who had taken at least one database course or had industry experience using database systems. 
The participants rated their experience in SQL, averaging at a score of 4.67 using a 7-point Likert scale (ranged from 3 to 6) with 1 being ``having no knowledge'' and 7 being ``expert'', 
which means most participants considered themselves non-expert database users.
None of them had used the graphical query builder before.
Each participant was compensated with a \$15 gift card.

\textbf{Data set.}
We used an academic publication data set used throughout this paper, which we collected from DBLP\footnote{\url{http://dblp.uni-trier.de/}} and ACM Digital Library.\footnote{\url{http://dl.acm.org/}}
It contains about 38,000 papers from 19 top conferences in the areas of databases (e.g., SIGMOD), data mining (e.g,. KDD), and human-computer interaction (e.g., CHI), since 2000. 
A relational schema was designed using standard design principles,
resulting in 7 relations with 7 foreign keys as depicted in Figure~\ref{fig:schema}.
As the main focus of this evaluation is on \name{}'s usability, this data set creates a sufficiently large and complex database for such purpose.

\textbf{Procedure.}
Our study followed a \textit{within-subjects design} with two conditions: the \name{} and Navicat conditions. 
Every participant first completed six tasks in one condition and then completed another six tasks in the remaining condition. 
The orders of the conditions were counterbalanced, resulting in 6 participants in each ordering.
We generated two matched sets of tasks (6 tasks in each set) differing only in their specific values used for parameters such as the title of the paper.
Before the participants  were given the tasks to carry out for each condition, they went through a 10-minute tutorial for the tool they would use. 
For each task, the participants could ask clarifying questions before starting, and they had a maximum of 5 minutes to complete each task.
After the study, they completed a questionnaire for subjective ratings and qualitative feedback.
Each study lasted for about 70 minutes.
Participants completed the study using Chrome browser, running on a Windows desktop machine, with a 24-inch monitor at a 1920x1200 resolution.

\textbf{Tasks.}
We carefully generated  two matched sets of 6 tasks
that cover many database exploration and querying tasks.
Table~\ref{table:tasks} shows one set (the other set is similar).
The tasks fall into three categories: 
finding attribute values (Tasks 1 \& 2);
filtering (Tasks 3 \& 4);
aggregation (Tasks 5 \& 6).
The tasks were designed based on prior research studies and their categorization of tasks.
Specifically, our categories are based on those used in database and HCI research \cite{amar2005low, li2014schema}, and our tasks vary in difficulty as in \cite{li2014constructing}.

\textbf{Measurements.}
We measured participants' task completion times.
If a participant failed to complete a task within 5 minutes, the experimenter stopped the participant 
and recorded 300 seconds as the task completion time.
After completing tasks for both conditions, the participants filled out a post-questionnaire that asked for their subjective ratings about \name{} (10 questions) and their subjective preference between two conditions (7 questions).

\begin{figure}[!tb]
  \begin{center}
  \includegraphics[width=\columnwidth,trim={2.7cm 6.0cm 2.4cm 5.7cm},clip]{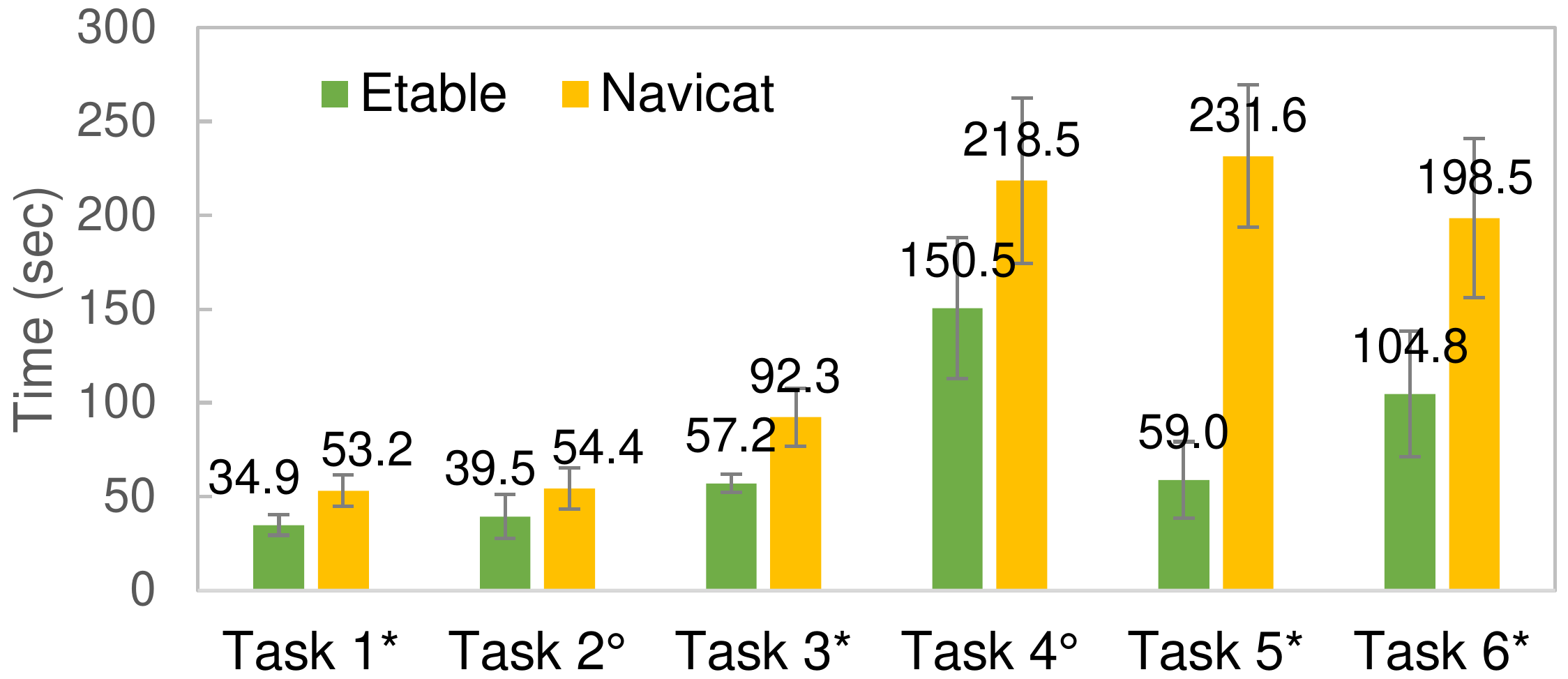}
  \vspace{-12pt}
  \caption{Average task completion time for each task. The error bars represent 95\% confidence intervals for the mean.
  Participants performed the tasks faster with \name{} than with Navicat. 
  The $^\ast$ and $^\circ$ symbols indicate 99\% and 90\% statistical significance in the two-tailed paired t-tests, respectively.}
  \label{fig:completion-time}
  \end{center} 
  \vspace{-10pt}
\end{figure}

\subsection{Results}

\textbf{Task completion times.}
The average task times for \name{} were faster than those for Navicat for all six tasks.
Figure~\ref{fig:completion-time} summarizes the task time results.
We performed two-tailed paired t-tests.
The differences were statistically significant for Tasks 1, 3, 5, and 6 
($p < 0.005$)
and marginally significant for Tasks 2 and 4 ($p = 0.052$, $p = 0.053$, respectively).
The results of Task 2 may be explained by an outlier participant who did not understand the requirement that each row of the final results must represent a different keyword.
Although Task 4 involves the highest number of operations that require participants to spend significant time in interpreting intermediate results before applying the next operators,
\name{} helped participants complete this task over 30\% faster than Navicat.

The task completion times for \name{} generally have low variance.
The larger variance in Navicat is mainly due to syntax errors that the participants faced.
Many participants, who are non-database experts, could not recall some SQL syntax and had trouble debugging errors.
In particular, they had trouble specifying GROUP BY queries in Navicat. For example, many participants did not specify a GROUP BY attribute in their SELECT clauses in their first attempts.
We also observed that many Navicat participants were overwhelmed by the complexity of the syntax of join queries~\cite{jagadish2007making} and preferred to specify new SQL queries from scratch instead of debugging existing ones when their original queries failed.
Unlike graphical query builders such as Navicat, \name{} helps nonexperts gradually build complex queries without having to know the exact query syntax.

\textbf{Subjective ratings.}
We asked participants to rate various aspects of \name{} using 7-point Likert scales (7 being ``strongly agreed'').
Their subjective ratings were generally very positive (see Table~\ref{table:ratings}).
In particular, all participants found \name{} easy to learn (i.e., rated 6 or 7), and almost all participants (11/12) found \name{} easy to use and helpful for browsing data in databases. 
They also enjoyed using \name{} (10/12) and would like to use software like \name{} in the future (11/12).
In response to the ``\textit{helpful to interpret and understand results}'' question, 
one participant commented that ``\textit{there are too many attributes ..., which is not easy to interpret.}''
To address this, as future work, we plan to develop techniques to rank and select the most important columns to show whenever a table has a large number of columns~\cite{yang2009summarizing}.

We also asked participants to compare \name{} and Navicat in 7 aspects.
All participants indicated that \name{} was easier to learn and was more helpful in browsing and exploring data.
A majority of participants liked \name{} more (11/12) and found it easier to use (10/12).
They would choose to use \name{} in the future (10/12) and felt more confident using it (8/12). 
Half of the participants answered that \name{} is more helpful in finding specific data than Navicat.
This result was expected because \name{}'s innovation focuses more on supporting data exploration.

\renewcommand{\arraystretch}{1.0}
\begin{table}[!tb]
\centering
\begin{tabular}{ l r  }
\toprule			
\textbf{Question} & \textbf{Avg.} \\
\midrule			
1. Easy to learn & 6.42 \\
2. Easy to use & 6.33 \\
3. Helpful to locate and find specific data & 6.25 \\ 
4. Helpful to browse data stored in databases & 6.67 \\ 
5. Helpful to interpret and understand results & 5.58 \\ 
6. Helpful to know what type of information exists & 6.00 \\ 
7. Helpful to perform complex tasks & 6.00 \\ 
8. Felt confident when using \name{} & 5.92 \\ 
9. Enjoyed using \name{} & 6.42 \\ 
10. Would like to use software like \name{} in the future & 6.50 \\
\bottomrule  
\end{tabular}
\vspace{-5pt}
\caption{Subjective ratings about \name{} using 7-point Likert scales (7: \textit{Strongly Agreed}. 1: \textit{Strongly Disagreed}).}
\label{table:ratings}
\end{table}

\textbf{Qualitative feedback.}
We asked participants about the features they liked about \name{}.
Many participants (9/12) explicitly mentioned the ``pivot'' feature.
They said that the pivot feature enabled them to easily specify complex join queries.
One participant said ``\textit{I also loved the pivot feature ... having multiple  pivots throughout the course of forming a query. I messed up a query, but could still find the right answer by doing an appropriate pivot.}'' 
In addition, many participants said that \name{} provides an intuitive view to users.
One said ``\textit{It is easy to see data from the perspective of what the users want to see/retrieve ...}''
Another said ``\textit{Visually, I was able to see ... the effects of the SQL operations, which made it easier to use and verify intermediate results.}''

\section{Expressiveness}
\label{sec:expressive}

This section discusses the expressiveness of the \name{} model.
We will first express the overall functionality of the  \name{} queries as a general SQL query pattern.
By doing so, we will show how typical join queries can be translated into \name{} queries, through multiple steps (similar to \cite{liu2009spreadsheet,catarci1997graphical}), demonstrating \name{}'s expressiveness.
Any join queries involving only FK-PK relationships on a relational database schema that meets \name{}'s assumptions (detailed in Appendix)
can be translated into an \name{} query that operates on TGDB.

The overall functionality of \name{} queries can be expressed as the following general SQL query pattern:
\begin{quote}
\vspace{-9pt}
\texttt{\\SELECT $\tau_a$.*, ent-list($t_1$),  ent-list($t_2$), $...$ \\
FROM $t_1, t_2, ...$\\
WHERE $source(p_1) = target(p_1)$ AND $source(p_2) =$\\       
\textcolor{white}{  } $\enspace \qquad target(p_2)$ AND $...$  AND $C_1$ AND $C_2$ AND $...$ \\
GROUP BY $\tau_a$; \\
}
\vspace{-9pt}
\end{quote}
where \texttt{ent-list} presents a list of corresponding entity references, similar to the \texttt{json\_agg} operator in PostgreSQL.\footnote{\url{http://www.postgresql.org/docs/9.4/static/functions-aggregate.html}}
Each of the four components  in an \name{} query (i.e., primary node type $\tau_a$, node types $T$, edge types $P$, and selection conditions $\mathcal{C}$) maps to a clause in SQL: 
primary node type to GROUP BY clause;
node\_types to FROM clause; 
edge\_types to join conditions; 
selection conditions to WHERE clause.

Following the above mappings, we now follow the approach similar to that in \cite{liu2009spreadsheet,catarci1997graphical} to show that \name{} can expressively handle typical join SQL queries, through a step-by-step translation.
That is, for any SQL join query following the above pattern, there exists an equivalent \name{} query.
\begin{enumerate}[noitemsep,topsep=0pt]
\item Transforms a relational algebra join expression ($R \bowtie R \bowtie ... $) to a graph relation correspondence $R^G * R^G * ...$ (described in Section 5.4)
by analyzing the list of relations in the FROM clause, and the join conditions in the WHERE clause.
(Each $R^G$ is a node type; each $*$ an edge type.)

\item Applies the original selection conditions  to the TGDB's node types;

\item If there is a \textit{group by} attribute, transform it to the graph's primary node type;
otherwise, if no \textit{group by} attribute exists, arbitrarily set a primary node type.
\end{enumerate}

\name{} can express typical join queries consisting of the core relational algebra (i.e., relational algebra expression that does not contain set operations),
which accounts for a large number of the database workloads.
\name{} 
additionally lets users choose a \textit{primary node type} from the list of selected relations, and introduces the \textit{entity-reference columns} (i.e.,  represented as \texttt{ent-list} in the above SQL pattern) to effectively present join queries.
This paper focuses on the critical usability challenge that arises when joining several tables.
In our future work, we plan to further increase \name{}'s expressiveness of the presentation model to the full set of operators in the relational algebra, 
through introducing additional operators to support more complex queries (e.g., set operations, complex aggregations, etc.).

\section{Conclusions}
\label{sec:conclude}

We proposed \name{}, a new presentation data model for interactively exploring relational databases.
The enriched table representation of \name{} generates a holistic, interactive view of databases that helps users browse relevant information at an entity-relationship level.
By directly interacting with the interface, users can iteratively specify operators,  enabling them to incrementally build complex queries and navigate databases.
\name{} outperformed a commercial graphical query builder in a user study, in both speed and subjective ratings across a range of database querying tasks.

This work takes a first step towards developing a practically usable, interactive interface for relational databases, and opens up many interesting opportunities.
Future research directions  include:
(1) incorporating more operations to further improve expressive power (e.g., set operations);
(2) accelerating the execution speed of updated queries 
(e.g., by reusing intermediate results);
(3) leveraging machine learning techniques to rank and select important columns to display.
The above ideas could usher a new generation of interactive database exploration tools that will benefit all database users.



\section{Acknowledgments}
This material is based upon work supported by the NSF Graduate Research Fellowship Program under Grant No. DGE-1148903
and the NSF under Grant No. IIS-1563816.

\bibliographystyle{abbrv}



\begin{appendix}

\section{Database Translation}
\label{sec:appendix}

This section describes a procedure for translating relational data\-bases into database schema and instance graphs in the typed graph model.
Our approach is based on the reverse engineering literature~\cite{batini1992conceptual, chiang1994reverse, catarci1997graphical, sequeda2012directly}.
We note that the following process cannot be applied to any relational schema, 
as relational schema do not contain all the semantics, 
but is a guideline for translations. 
We make several assumptions as in the literature~\cite{catarci1997graphical, batini1992conceptual}.
First, all the relations are in BCNF or 3NF.
Second, there are no ternary relationships: all the relationships are binary.
Third, for relationship relations, we assume that all attributes are foreign keys of the relations that participate in the relationship. Any attributes of the relationship itself are ignored.
Finally, a relation representing a multivalued attribute always consists of two columns.

\noindent \textbf{Identifying entity relations.}
This step identifies entity relations from a set of relations. Informally, entity relations refer to relations constructed from entity types in the entity-relationship model. We define an \textit{entity relation} as a relation whose primary key does not contain a foreign key or a key inclusion dependent on any other attribute in any other relation~\cite{catarci1997graphical, batini1992conceptual}.
For each of the identified entity relations, the following process is performed.
\begin{enumerate}[noitemsep,topsep=0pt]
\item A relation becomes a node type  in the schema graph.
\item The relation name becomes the name of the node type. 
\item All  the attributes of the relation become the attributes of the node type.
\item One attribute selected by users becomes a \textit{label attribute} for the node type.
\end{enumerate}

We currently determine the \textit{label attribute} based on a combination of heuristics, such as
data type (e.g., text generally more interpretable than numbers) and cardinality.
However, this label selection task is hard to fully automate.
Thus, we also allow users to manually pick a desired label attribute.
In our future work, we plan to investigate mixed-initiative approaches that allow human and computer to work together, so that we would provide an initial guess and recommend possible alternatives based on the heuristics, and allow the users to select attributes that are most  meaningful to them.

\noindent \textbf{Identifying 1:1 and 1:n relationships.}
Foreign keys, which are used to represent one-to-one and one-to-many relationships between entity relations in the relational model, are used to identify relationships between entity relations found above.
For each foreign key, the following process is performed.
\begin{enumerate}[noitemsep,topsep=0pt]
\item Each foreign key becomes an edge type in the schema graph. The source node would be a node type representing a relation containing the foreign key. The target node would be a node type representing a relation which the foreign key refers to. 
\item Unless the source and target node types are the same, the edge types are duplicated with a reverse direction.
\item The label is defined as the name of the target node type. If the label is used by another edge type, a slightly different label will be created.
\end{enumerate}

\noindent \textbf{Identifying many-to-many relationships.}
Many-to-many relationships are represented as a separate table in the relational model. We identify these tables whose primary key is a concatenation of primary keys of two other entity relations.
For each of the identified relationship relations, the following process is performed.
\begin{enumerate}[noitemsep,topsep=0pt]
\item Each relationship relation becomes an edge type in the schema graph. The two other associated entity relations become source and target nodes. 
\item The remaining steps are the same as above (i.e., Steps 2 \& 3) 
\end{enumerate}

\noindent \textbf{Identifying multivalued attributes.}
The relational model stores multivalued attributes in separate relations. We identify such relations. We assume these relations consist of only two attributes where both attributes make up the primary key and the first attribute is a foreign key to an entity relation.
For each of this case, the following process is performed.
\begin{enumerate}[noitemsep,topsep=0pt]
\item The second attribute becomes a node type in the schema graph. 
\item The node type has one attribute which refers to itself. The label column is this only attribute.
\item An edge type is also created from the node type representing the entity relation to the newly created node type. It will be duplicated in a reverse direction.
\end{enumerate}

\noindent \textbf{Identifying categorical attributes.}
This step of identifying categorical attributes is optional, but we find it useful.
People often perform GROUP BY operations over categorical attributes, and this step helps them perform such analysis.
Any of the attributes of the entity relations could be selected by users. Often, attributes with low cardinality (e.g., less than 30) can be candidates for categorical attributes. For each of the selected attributes, the following process is performed.
\begin{enumerate}[noitemsep,topsep=0pt]
\item Each attribute becomes a node type in the schema graph. 
\item It has one attribute which refers to itself. The label column is this only attribute.
\item An edge type is also created from the node type representing the relation to the newly created node type. It will be duplicated in a reverse direction.
\end{enumerate}

This creates a TGDB schema graph. 
Under the assumptions we made, the schema graph contains all the information in the original relational schema. 
Once the schema is translated, it is straightforward to create the corresponding TGDB instance graph.

\end{appendix}

\end{document}